\documentstyle[12pt,aasms4]{article}

\received{..}
\accepted{..}
\journalid{The Astrophysical Journal}{}

\lefthead{Moll\'{a},Hardy,Beauchamp}
\righthead{The Stellar Populations in Spiral Disks}

\def\msun{\thinspace\hbox{$\hbox{M}_{\odot}$}}

\begin{document}
\setcounter{page}{1}
\def\counter{1}

\title{THE STELLAR POPULATIONS OF SPIRAL DISKS.II. MEASURING AND MODELING 
THE RADIAL DISTRIBUTION OF ABSORPTION SPECTRAL INDICES}

\author{MERCEDES MOLL\'A}
\affil{D\'epartement de Physique, Observatoire du mont M\'egantic, 
Universit\'e Laval, Sainte-Foy, P.Q., Canada, G1K 7P4}
\author{EDUARDO HARDY$^1$}
\affil{National Radio Astronomy Observatory, Casilla 36-D, Santiago, Chile}
\author{DOMINIQUE BEAUCHAMP}
\affil{D\'epartement de Physique, Observatoire du mont M\'egantic, 
Universit\'e Laval, Sainte-Foy, P.Q., Canada, G1K 7P4}
\altaffiltext{1} {Visiting astronomer,
Cerro Tololo Inter-American Observatory, National Optical Astronomy 
Observatories, which are operated by the Association of Universities 
for Research in Astronomy, Inc. under contract with the National Science 
Foundation.}



\today


\begin{abstract}

The radial distributions of the Mg$_{2}$ and Fe5270 Lick spectral
indices have been measured to large radial distances on the disks of
NGC~ 4303 and NGC~ 4535 using an imaging technique based on
interference filters. These data, added to those of NGC~ 4321
previously published in Paper I of this series are used to constraint
chemical (multiphase) evolutionary models for these galaxies. Because
the integrated light of a stellar disk is a time average over the
history of the galaxy weighted by the star formation rate, these
constraints complement the information on chemical gradients provided
by the study of H\,{\sc ii} regions which, by themselves, can only
provide the $\alpha$--elements abundance acummulated over the life of
the galaxy.  The agreement between the observations and the model
predictions shown here lends confidence to the models which are then
used to describe the time evolution of galaxy parameters such as star
formation rates, chemical gradients, and gradients in the mean age of
the stellar population.

\end{abstract}

\section{INTRODUCTION}

Abundance radial gradients in the Milky Way Galaxy (MWG) have been
widely studied (see e.g. Prantzos \& Aubert 1995) and a large number
of models to reproduce and interpret them have been developed starting
with that of Tinsley (1980).  In the latter the Galaxy chemical
evolution was described by the simplest model of a closed box, which
failed to reproduce the large observed radial gradient.  Different
mechanisms were suggested afterwards to generate these radial
gradients ( G\"{o}tz \& K\"{o}ppen 1992).  Gas infall remains the most
often invoked mechanism capable of reproducing the observations of the
MWG. This mechanism can also solve the well known G-dwarf problem
(Pagel 1989, Pitts \& Tayler 1989).  Several numerical models
incorporating gas infall have been used to describe the MWG (see Tosi,
1996 and references therein) with various degrees of success. The
multiphase model described in section 3 of this paper is one of them.

Variations of abundances with galactocentric radius have also been
observed in other spiral galaxies (D\'{\i}az 1989; Skillman et
al. 1996; Garnett et al. 1997; Henry 1998). Abundance radial gradients
give different dex/kpc slopes for different galaxies and some
correlations between these values and galaxy characteristics have been
shown to exist (Vila-Costas \& Edmunds 1992; Zaritsky, Kennicutt, \&
Huchra 1994).

Until now abundance radial distributions in spiral galaxies have been
 studied only with data pertaining to the gas phase where abundance
 estimations are obtained from observations of H\,{\sc ii} region
 emission lines. Only for our Galaxy (Edvardsson et al. 1993;
 Edvardsson 1998) and for some local group galaxies, such as M33
 (Monteverde et al. 1997, there are abundance data belonging to the
 stellar component of disks.  Therefore, only the present-day
 abundances are known in external spiral galaxies, and they refer only
 to the $\alpha$ group of elements such as oxygen, calcium or sulfur,
 since the iron group is not accessible to H\,{\sc ii} region
 observations.

For gas-depleted objects such as elliptical galaxies, bulges and
globular clusters, the study of integrated stellar spectral indices
has been the most common technique used to estimate abundance
levels. In particular the Lick indices Mg$_{2}$ and Fe5270 have been
widely used and a large number of observations have accumulated (Faber
1973; Mould 1978; Burstein 1979; Burstein et al. 1984, Faber et
al. 1985; Baum, Thompsen, \& Morgan 1986; Davies et al. 1987; Couture
\& Hardy 1988; Worthey, Faber, \& Gonz\'{a}lez 1992; Davidge 1992;
Jablonka, Martin, \& Arimoto 1996; Fisher, Franx, \& Illingworth 1996;
Trager et al.  1997). In these cases the interpretation of the above
indices, although far from trivial, is considerably simplified by the
probable narrow age interval of the stellar population under
study. Furthermore, these are generally high surface brightness
objects.  Due to their lower surface brightness, however, similar
observations for the stellar component of spiral disks are extremely
difficult to achieve, even with large telescopes. The absence of this
information is regrettable because the integrated light of spiral
disks contains a light-weighted, age-weighted chemical signature
complementary to that of H\,{\sc ii} regions.

In the present work we use a new imaging technique to obtain
information on the Mg$_{2}$ and Fe5270 spectral features across the
face of spiral disks (Beauchamp \& Hardy 1997, hereafter Paper I).
Thus, we are observing abundance indicators averaged over the chemical
history of the disk as opposed to the accumulated effect on the
extreme population I component revealed by the H\,{\sc ii} regions or
by B stars.  But spectral indices of composite systems do not measure
abundances directly, and to derive the latter from the former require
the help of evolutionary models. We will discuss in this paper the use
of a multiphase chemical evolution model for this purpose.

The multiphase chemical evolution model has been recently applied to
six spiral galaxies of different morphological types (Moll\'a,
Ferrini, \& D\'{\i}az 1996, hereafter MFD96).  This model may be able
to explain the observed correlations of abundance gradients with
galaxy types as arising from variations of the characteristics infall
rate and of the cloud and star formation efficiencies with
morphological type and/or Arm Class. The observed radial distributions
of diffuse and molecular gas, oxygen abundances and star formation
rate can in turn be used as constraints for these models.

The time evolution of abundance radial gradient, and other quantities
involved in chemical evolution, such as the star formation rate (SFR)
or gas and stellar surface densities, may also be obtained from the
numerical theoretical models. These time-dependent quantities have
been calculated with the multiphase model for the same group of six
spiral galaxies (Moll\'{a}, Ferrini, \& D\'{\i}az 1997, hereafter
MFD97).  However, when only observations of H\,{\sc ii} regions are
available, it is only the present day abundance that serves as a
constraint for the model. As a result, the time evolution obtained
with our models is not well constrained and different evolutionary
paths may lead to the same final outcome.  In fact, models for the MWG
using similar present-time radial abundance distributions yield
different histories of the abundances gradients (Moll\'{a}, D\'{\i}az,
\& Ferrini 1992; Tosi 1996) as a result of differences in SFR and
infall rate behavior in time and space.  Fixing these physical
parameters would require information from intermediate evolutionary
phases.

A way to constrain more tightly this time evolution will be proposed
here.  The aim is to compare the predicted radial distribution of
stellar absorption spectral indices, calculated by assuming a given
stellar abundance in the disk, with the observed distribution of
spectral indices. The latter represents a time-averaged quantity over
the life of the disk strongly weighted by its SFR.

The plan of this paper is as follows. In section 2 we summarize our
observational technique and the data obtained for the three galaxies,
NGC~ 4303, NGC~ 4321 and NGC~ 4535. We present the description and
physical input of our numerical models in Section 3.  In Section 4 we
combine our chemical evolution calculations with synthesis results for
Single Stellar Populations (SSP) in order to predict radial
distributions of the two spectral indices, Mg$_{2}$ and Fe5270 adopted
for this investigation, and then compare them with the
observations. In Section 5, we extract and discuss the resulting
information on the characteristics of the stellar populations present.
Finally, we present our conclusions in Section 6.

\section{OBSERVATIONS: TECHNIQUES AND DATA}

In this section we summarize the observational techniques used in this
work to measure radial metallicity profiles in the inter-arm regions
of the discs of spiral galaxies. For a more exhaustive description of
the technique we refer the reader to Paper I (Beauchamp \& Hardy
1997).

\subsection{The Technique}

The basic idea of the technique described here is to make use of
spectral images to increase the signal to noise of the index
measurement by adding up azimuthally all the information contained
within a given radial interval. By comparison, traditional long-slit
spectroscopy provides multiplex spectral information but only over a
very restricted surface. The advantage of our technique is that we are
able to follow a restricted selected spectral interval over a large
spatial interval with excellent signal to noise ratios using small
telescopes.  Azimuthal integration of the underlying smooth stellar
signal, after removal of the signature of the spiral arms and
associated extreme pop. I structures, provides measurements of
spectral indices useful to radial distances where the surface
brightness of the galaxy reaches $\sim$ 24$\mu_V$ .

The first step followed consisted in obtaining monochromatic images of
galaxies in order to measure the Lick Mg$_2$ and Fe52 indices.  We
used four intermediate pass band (60\AA ) filters, two of them
centered on the magnesium and iron features and two on continuum
regions, RC for the red and BC for the blue continua.  A description
of the filters is given in Table 1. The interference filter set
described there was designed for a mean velocity close to that of the
Virgo cluster. At that distance we have excellent spatial resolution
per flux unit, atmospheric seeing does not become an important
problem, and our bandpasses do not become contaminated by any strong
sky emission line. We designed our filters to closely reproduce the
Lick definition (see Faber et al. 1985) for Mg$_2$, but we chose to
use the same continua for Fe52 than for Mg$_2$ in order to save
observing time. We thus obtained a pseudo-index, Fe52', that we
calibrated later into the standard Lick Fe52 system at the expense of
some loss of sensitivity. The adopted band-passes are shown
graphically on Fig. 1 of Paper I.

For each galaxy we obtained a set of four images from where we could
compute the Mg$_2$ and Fe52 indices at almost any point in the
galaxy. We must emphasize that we choose galaxies that were nearly at
the same redshift because our filters could not be significantly tuned
(tilting allowed a range of only about 5\AA). In addition we selected
galaxies viewed nearly face-on in order to increase the number of
available pixels, better isolate the inter-arm pixels, and reduce
rotational effects.

 Given the widths of the filters (about 60 ${\rm \AA\:}$ FWHM) which
were almost rectangular, the average nightly temperature, and the
tilting capabilities, we could observe galaxies spanning the
approximate velocity interval 1400~${\rm km}\:{\rm s}^{-1}$ --
2100~${\rm km}\:{\rm s}^{-1}$. This interval was quite appropriate for
Virgo spirals. Notice that the strong ${\rm [OII] \: \lambda 5577 \AA\
} $ atmospheric emission line falls to the red of our filters, an
important consideration when dealing with photometry at small surface
brightness levels (see Baum, Thompsen, \& Morgan 1986).

The second observational step involved the computation of the zero
point of the index system, which depends on atmospheric conditions and
thus must be obtained from multiplexing spectroscopy of the bright
nuclear regions of the target galaxies. The spectroscopy was
calibrated against the standard Lick system, and the
spectroscopically-derived Lick indices were compared with the ones
obtained for the same region from the interference-filter images, thus
yielding the required zero points.

Using the appropriated color and brightness criteria V, R and I images
were then used to reject the young and hot population I lying in the
spiral arms.

\subsection{The Data}

The image data set was obtained during two observation runs, one at
CTIO (Cerro Tololo Inter-American Observatory) during March 1995 for
NGC~ 4535 and at OMM (Observatoire du mont M\'egantic) in March 1996
for NGC~ 4303 and NGC~ 4321. At CTIO the 90cm telescope was used with
a 2k x 2k CCD binned to a 1k x 1k format. The pixel size of
0.79$\arcsec$ gave a field of view of 13.5$\arcmin$. At OMM the 1.6m
telescope was used in conjunction with a 1k x 1k CCD and the
"Panoramix" focal reducer.  For this one, the pixels' size of
1.18$\arcsec$ gave a field of view of 20$\arcmin$. At CTIO, a spatial
resolution FWHM slightly below 2 arc-seconds was reached while it was
between 3$\arcsec$ and 4$\arcsec$ at OMM.

The spectra were obtained with the OMM Boller \& Chivens long slit
spectrograph during five runs conducted from January 1995 to June
1996.  In addition to the bright central portions of the galaxies we
took spectra of standard stars in order to place the data on the Lick
system. The reader is referred back to paper I for details of the
observational setups used on both Observatories.

\section{CHEMICAL EVOLUTION MODELS}

We applied the multiphase model described in detail in MFD96 to the
galaxies NGC~ 4303, NGC~ 4321 and NGC~ 4535. These three galaxies are
barred and are located in the Virgo cluster.  Their morphological
types (T) and Arm Class are shown in Table 2, columns (2) and (3), and
were taken from Tully (1988) and Biviano et al. (1991),
respectively. Distances in column (4) and effective radii in column
(5) were also taken from Tully (1988) with the exception of NGC ~4303,
for which the effective radii was taken from Henry et al. 1992.

The chemical evolution for these 3 galaxies was modeled as described
in MFD96. We assumed that every protogalaxy was initially a spheroid
composed of primordial gas. The total mass included in the spheroid,
as well as its radial distribution, are derived from the H\,{\sc i}
rotation curves taken in this case from Guhathakurta et
al. (1988). The maximum rotation velocity (in km/s) given for these
authors for each galaxy is in column (6) of Table 2.

Optical (H${\alpha}$) rotation curves by Chincarini \& de Souza (1985)
for NGC~ 4535, and Distefano et al.  (1990) for NGC~ 4303 and NGC~
4321 have also been used.  These curves cover a short radial range,
but they map better the inner regions.  As a result we tried to use
both types of curves. This was easy for NGC~ 4321 and NGC~ 4535, but
proved to be more cumbersome for NGC~ 4303 for which both curves were
quite different in the 1--5 kpc range. Thus, the total mass and its
radial distributions are better determinated for NGC~ 4321 and NGC~
4535 than for NGC ~ 4303. In particular, if we take into account the
larger rotation velocity estimated with optical data for the inner
regions of the latter galaxy, a higher total mass at 2--3 kpc
galactocentric distance is found. Moreover, as pointed out, these
three galaxies have a bar. We do not know if these rotation curves
trace the implied accumulation of gas or stars in the bar region.

We now follow the protogalaxy evolution. Initially all of the mass is
in the form of diffuse gas. This gas begins to form stars in the halo
while collapsing onto an equatorial plane from where the disc will
form.  The model splits the galaxy into concentric cylindric regions 1
kpc wide. There is a {\sl halo} zone and a {\sl disk} one at each
galactocentric distance. The collapse proceeds at a rate that is
inversely proportional to the collapse time scale. This time scale
$\tau (R)$ in turn depends on the galactocentric distance due to its
dependence on the total mass surface density: $\tau \propto \sigma
(R)^{-1/2}$ (Jones \& Wyse 1983). We assume the resulting radial
dependence to be an exponential function, namely $\tau (R)=\tau _0
e^{{-R/l}}$, where $l$ is the scale length taken as 4 kpc for the
three galaxies. The characteristic collapse time scale $\tau _0$
depends on the total mass of the galaxy through the relation $\tau
_0\propto M_9^{-1/2}T$ (Gallagher, Hunter, \& Tutukov 1984), where
$M_9$ is the total mass of the galaxy in 10$^9\msun$ units and $T$ is
its age, assumed to be 13 Gyr for all spiral galaxies. We calculate
the collapse time scale $\tau _0$ for the program galaxies via the
expression $\tau_{0}=\tau_{\odot}(M_{9,gal}/M_{9,MWG})^{-1/2} $, where
the subscript MWG refers to the Milky Way, and $\tau_{\odot}$, taken
as 4 Gyr, is the collapse time scale value used for the solar
neigbourhood region (Ferrini et al. 1992), which we use as a
reference.  Total masses $ M_{9,gal}$ are calculated from the rotation
curves via the expression $ M_{9,gal}= 2.32\times 10^{5} R_{gal}
V_{max}^{2}$ (Lequeux 1983), with $R_{gal}= 30$ kpc for every galaxy
whereas $V_{max}$ takes the values listed in Table 2. This collapse
time is associated for every galaxy with a region equivalent to the
solar region in the MWG, located at a galactocentric distance $R_0$
which is itself calculated from the individual effective radius
$R_{eff}$ (see table 2, column 5).  Values for $R_{0}$ and $\tau (
R_{0}) = \tau_{0}$ are listed in Table 3, where the model parameters
are shown (see below).

We now let stars form and allow them to evolve and enrich the ISM. The
initial mass function (IMF), an empirical function similar to a Scalo
law, is taken from Ferrini, Palla, \& Penco (1990). The required
nucleosynthesis prescriptions for massive stars are those of Woosley
\& Weaver (1986). In the specific case of $^{14}$N production the
prescriptions come from Maeder (1983). Single low and intermediate
mass stars contribute to the enrichment through winds and planetary
nebulae by producing N, C and He following Renzini \& Voli (1981)
prescriptions.  The type I supernovae explosions release mostly iron
following Nomoto, Thielemann, \& Yokoi (1984) and Branch \& Nomoto
(1986).

The star formation in the halo is assumed to follow a Schmidt law with
an exponent of the gas density $n=1.5$. The proportionality coefficient
depends on the radial distance of the halo region being considered and
on the efficiency of the process, $\epsilon_{K}$, which is assumed
constant for every region and galaxy.

Star formation in the disk is assumed to be a two--step process:
first, molecular clouds form from the diffuse gas; then spontaneous
star formation is triggered by cloud-cloud collisions. Stimulated star
formation by the interaction of massive stars with molecular clouds is
also included in the computation.  Because star formation in molecular
clouds is a well-observed process, our star formation prescription is
probably closer to physical reality than other frequently assumed
laws.  Each one of formation rates used throughout the computation
depends on the masses of the phases involved with proportionality
factors that depend in turn on efficiencies or probability factors.
The actual equations describing these parameters are obtained in
Ferrini et al. (1994), where their radial dependence are calculated.
We direct the reader to this reference for a thorough description of
the multiphase chemical evolution model and parameters involved.
Induced star formation is a local process with an efficiency $\epsilon
_a$ constant for all regions and galaxies. Cloud and stimulated star
formations have a radial dependence as a result of volume effects, and
a dependence on the efficiencies, namely $\epsilon _\mu $ for the
former and $\epsilon _H$ for the latter process.  The acceptable range
of these efficiencies is determined by taking into account the Hubble
type, and/or the Arm Class for every galaxy. Galaxies of early types
of Hubble and/or Arm Class higher are assumed to have larger values of
$\epsilon _\mu$ and $\epsilon _H$ than the later type ones, or lower
Arm Class due to the stronger density wave and tighter arms.  The
latter physical properties can compress better the diffuse gas phase
by favoring the formation of molecular clouds and by increasing the
frequency of cloud-cloud collisions. These arguments, developed by
Ferrini \& Galli (1988) were applied to previous models. In particular
a discussion of the possible values of these efficiencies for
different types of galaxies and a comparison with observed values are
given in MFD96.  Taking into account these arguments the galaxy NGC~
4321, with an Arm Class of 12, must have larger efficiencies than NGC~
4303 with an Arm Class is 9, although the Hubble type is the
same. NGC~ 4535 has both T and the Arm Class very similar to those of
NGC~ 4303; therefore their efficiencies must be similar, too.  These
efficiency values are applied within a region similar to the solar
neighborhood located at a radius $R_{0}$.

The input parameter values, i.e., final efficiencies and collapse
times have been adjusted by fitting the model output to the observed
radial distributions of atomic gas (H\,{\sc i}), molecular gas (H$_2$)
and star formation rate surface densities, and for the oxygen
abundance of the gas phase.  Observational data for H\,{\sc i} surface
densities are from Warmels (1988) and Cayatte et al. (1990).
Molecular gas density radial distributions are obtained from Kenney \&
Young (1988,1989) after suitable corrections needed to take into
account the radial dependence of the factor $\chi $ that transforms
intensities $I_{CO}$ into molecular gas densities. Following Verter \&
Hodge (1995) and Wilson (1995), this factor depends on variations in
the abundances of the gas, decreasing below its solar value when
abundances are larger than solar and increasing when abundances are
sub-solar.  The star formation radial distribution is obtained from
H$\alpha $ fluxes of Kennicutt (1989) as normalized to the
``equivalent--to--solar'' region value $\Psi_{0}$.  For the galaxy
NGC~ 4321 the most recent data from Knapen \& Beckman (1996) and
Knapen et al.  (1996), kindly made available to us by the first
author, has been used.  These data set consists of radial
distributions of atomic and molecular gas densities and star formation
rate derived from the H$_{\alpha}$ fluxes.  Oxygen abundances are
taken from Henry et al. (1992) for NGC~ 4303, and from McCall, Rybsky,
\& Shields (1985) and Shields, Skillman, \& Kennicutt (1991) for NGC~
4321.  There are no observed elemental abundance data for NGC~ 4535.
In the first two cases the calibration of Edmunds \& Pagel (1984)
between $R_{23}$ and oxygen abundance $12+log(O/H)$ has been used.
Whether other parameters are also able to fit these present gas radial
distribution and the radial distribution on Mg$_{2}$ and Fe52 spectral
indices is an issue that will be analyzed in a forthcoming paper
(Moll\'{a} et al. 1998), where the issue of sensitivity to parameter
variations, and thus the uniqueness of this set of efficiencies will
be discussed.

The final tuning of efficiencies and collapse times yields values
similar but somewhat lower than those chosen in previous models for
field galaxies of the same luminosities and morphological types. This
is consistent with our program spirals behaving as if they actually
were of later morphological types than hitherto classified. This
suggestion is in agreement with the recent work by Koopmann \& Kenney
(1998) where, by comparing the Hubble type classification for field
and cluster galaxies, they find that cluster galaxies seem to be
classified as earlier than isolated field galaxies having the same
absolute bulge strength.

We show in Figs. 1--3 our results for chemical evolution models
characterized by the input data of Table 3.  These model results are
shown by the solid line in figures 1, 2 and 3 for NGC~ 4303, NGC~ 4321
and NGC~ 4535, respectively, together with the corresponding
observational data. These figures display abundance radial
distributions in panels (a), surface densities of star formation rate
as normalized to the value corresponding to the equivalent solar
region $\Psi_{0}$ in panels (b) and finally, gas radial distributions
for both the difusse gas H\,{\sc i} (panel c) and for the molecular
gas H$_{2}$ (panel d).

As in previous work, diffuse gas radial distributions results are very
well fitted by the models for our three galaxies, showing that the
efficiency values as well as their radial dependences,
$\epsilon_{\mu}( R)$, \footnote{We assume, as in previous multiphase
models, that the effect of the density wave is larger in the inner
regions of disks by adopting increasing efficiencies $\epsilon_{\mu}$
for decreasing radii} for those processes that transforms the diffuse
gas into molecular gas are realistic.  The model-predicted star
formation and molecular gas radial distributions of previous works
(MFD96) were lower than observed in the inner regions. This points
towards a problem with the selection of efficiencies or with the
assumed radial dependence for the destruction of molecular clouds and
their conversion into stars, at least for the inner regions of disks.
It is likely that this inconsistency was related to the treatment in
our previous models of the threshold density for the beginning of the
star formation process.  In effect, observations seem to require a
minimal value of the gas density in order for star formation to be
triggered. Following Kenniccutt (1989) this density value depends on
the velocity dispersion of the gas.  This velocity dispersion is
almost constant along the disk galactocentric radius in all galaxies
and, because of this, the star formation depends directly on the
rotation velocity and inversely on the radius. The latter dependence
is already included in our model via the volume dependence of
parameters for cloud and star formation processes. Thus, the {\sl
standard} multiphase model star formation law already simulates the
behavior that would take place with the inclusion of a threshold
density for the triggering of the star formation process. The previous
assumption of a constant efficiency $\epsilon_{H}$ for spontaneous
star formation from cloud-cloud collisions is reasonable provided that
the velocity dispersion remains constant. However, in the inner
region, close to the bulge, where the disk thickness is larger, this
velocity dispersion is larger than elsewhere in the disk (Bertola,
Cinzano, \& Corsini 1995; Palacios et al. 1997), and the resulting
effect is to increase the threshold density to form stars. We have
simulated this effect by introducing a variable star formation
efficiency for cloud-cloud collision that decreases with decreasing
radius.  With this assumption the radial distributions of molecular
gas are much better reproduced and those of the star formation surface
densities are reasonably well fitted. The exception is however the
galaxy NGC~ 4321.  In the latter case the difference between the
predicted and observed star formation radial distribution could
perhaps be explained by the presence of a bar.

In every case studied we have detected a problem with the radial
distribution of oxygen abundance in that we have found impossible for
our models to predict values as large as those observed in the inner
regions. A factor that may explain these high abundances near the
center is the existence of a bar. Following Friedli \& Benz (1995)
models, a bar should produce a radial flow of material towards the
center thus flattening the radial gradient, but in the region where
this gas flow produces a starburst the abundance should actually
increase.  This phenomena is local and lasts a few Myr leading to a
new maximum in the oxygen abundance. Clearly this abundance effects
cannot be reproduced by multiphase models without the introduction of
radial flows.

 However, $ 12 +log (O/H)$ abundances larger than 9.1/9.2 cannot be
obtained at all by the multiphase chemical evolution models, which
reach saturation level for the oxygen abundance.  This means that
observed abundances larger than 2 times solar are never reproduced, an
old and well-known problem for theoretical chemical evolution
models. Maybe the nucleosynthesis prescriptions are not adequate
(usually their variations with metallicity are not taken into account)
or the observed abundances as obtained via uncertain calibration
methods are overestimated (see D\'{\i}az 1998).

\section{USING SYNTHESIS MODELS TO DESCRIBE GALAXY EVOLUTION}

As a byproduct of the model applied here we can derive, as we show
 below, the time evolution of every region for which we are able to
 reproduce the observed characteristics. We can find for example the
 star formation history and the time evolution of the various
 abundances.  With both pieces of information we can in turn calculate
 the integrated mass of stars created in a given time step and the
 mean abundance reached at that point in time by the gas from which
 stars form. These values are associated to a mean time in order to
 calculate a mean age for each stellar generation. The time step used
 in our calculations is 0.05 Gyr.  Therefore the youngest stellar
 population may be 5$\times$10$\rm ^{7}$ yr old.  If we were to use
 smaller time step, the computation time would increases unduly, but
 we have found that model results are not too sensitive to the size of
 the time step.

What all of the above means is that we may consider the stellar
 populations present at each galactocentric region as a superposition
 of single stellar populations or {\sl generations} each characterized
 by its age and metallicity. Moreover, our model yields abundances for
 15 elements, including oxygen, magnesium and iron, so we know the
 ratio [Mg/Fe] for every one of these generations. The latter is a
 very important datum, being a time--dependent indicator of the mix of
 $\alpha$--elements and Fe--peak elements that tracks the formation
 history of the galaxy.

We now describe how to use the observed spectral indices as
constraints to our evolutionary models. We emphasize that spectral
indices do not measure abundances, but must be transformed into them
via the computation of stellar populations of known ages and
metallicities. These populations can be built by combining spectra
from stellar libraries containing a wide span of physical
parameters. Thus, each spectral index can be empirically correlated
with stellar gravity, effective temperature and metallicity, providing
in this way a set of `Fitting Functions' covering the stellar
parameter space. In the particular case of the two indices adopted
here a significant effort have already been made in this direction
(Gorgas et al. 1993; Barbuy 1994; Worthey et al. 1994; Idiart \&
Freitas-Pacheco 1995; Borges et al. 1995- hereafter B95). The last two
works cited used measures of the indices Fe52 and Mg$_{2}$ derived
from their own sample of stars providing the direct dependence of Mg
index on the ratio [Mg/Fe].  Synthesis models giving spectral indices
for single stellar populations (SSP) of a given age and metallicity
have also been calculated (i.e. , Brodie \& Huchra 1990; Buzzoni,
Gariboldi, \& Mantegazza 1992, Buzzoni, Mantegazza, \& Gariboldi 1994;
Worthey (1994), B95, Weis, Peletier, \& Matteucci 1995; Casuso et
al. 1996; Vazdekis et al. 1996), all of them with the study of
elliptical galaxies in mind.  We use here the B95 results because
these authors provide the ratio [Mg/Fe] explicitly.

We assume that each generation of stars can be represented by a given
SSP of known physical parameters. Each spectral index is then
calculated from B95 (their equations (9) and (11)), using as
parameters the iron abundance or metallicity [Fe/H], the ratio
[Mg/Fe], and the age of the generation as provided by the galaxy
models.  In addition we assign a total continuum flux to each spectral
feature using the explicit SSP spectra from Worthey (1994). With these
continuum fluxes and indices we can then compute the fluxes under both
spectral features. Finally, we weight these results by the stellar
mass created on each time step as provided by the best-fitting
multiphase model, using for this purpose Worthey's (1994)
mass/luminosity ratio corresponding to each generation.  This
calculation is performed for all regions of the disk in order to
obtain predicted spectral indices along its radius.  The resulting
radial distributions of predicted spectral indices are represented by
the solid lines in Figures 4, 5, and 6.  Notice the procedure followed
here to relate spectral indices to abundances: since the
index--to--abundance transformation requires knowledge of the
population mix present in the disk, we obtain the mix from the model
and then use SSP models for every generation in the mix to derived the
integrated Lick indices which we then compare to the observations.

In Figures 4,5 and 6 we show the predicted radial distributions of
Mg$_{2}$ (panels a) and Fe52 (panels b) for NGC~ 4303, NGC~ 4321 and
NGC~ 4535 together with the observational data.  This is only a first
order model which does not follow the local features of disks and
attempts only to reproduce the azimuthally-averaged abundance
gradient. Keeping this in mind, the observational data are
sufficiently well reproduced in all cases by the model calculations as
to lend confidence to the models.  In the following section we will
use the model output to describe the stellar populations of each of
the observed disk.

\section{RESULTS AND DISCUSSION}

Having lent credibility to the models via the use of the integrated
Lick indices, as described in the previous section, we now use the
model output to follow the temporal and spatial evolution of the
galaxies.

In Figures 7, 8 and 9 we display the time evolution of abundance
gradients for the multiphase models of the three galaxies under study.
In panels (a) we show the $ 12 +log (X/H)$ abundances of magnesium
(solid line) and iron(dashed line) for three different regions
corresponding to the inner, the central and the outer disk. Inspection
of these figures reveals some important facts about the abundance
gradients and levels, which we presently discuss.  The magnesium and
the oxygen, being $\alpha$-elements, are very quickly produced by
massive stars and the abundance gradient is rapidly flattened because
abundances reach the saturation level via gas depletion in a short
time. The radial gradient of [O/H], for the galaxy NGC~ 4303,
decreases from a value -0.141 dex kpc$^{-1}$ at an age of 2.5 Gyr to
-0.105 dex kpc$^{-1}$ at 5 Gyr to finally reach a present value of
-0.05 dex kpc$^{-1}$.  For the magnesium these gradient values are
similar, being -0.143, -0.107 and -0.053 dex/kpc, respectively.  Iron,
on the other hand, is produced by type I supernova explosions from low
mass stars, and as a result it is ejected much later. Thus the radial
gradient of [Fe/H] remains steeper for a longer time: -0.148 dex/kpc
at 2.5 Gyr, -0.124 at 5 Gyr and -0.07 at 13 Gyr.

As for the abundance levels, gas abundances appear to be less
time-dependent for the inner than for the outer regions. This in turn
suggests that stellar populations of different ages may have very
nearly the same abundances in the inner regions, and a larger spread
in the outer ones.

As expected, the relative abundance ratio [Mg/Fe] starts with a high
value and decreases in time as the iron appears. The rate of decrease
of [Mg/Fe] depends on the star formation rate. Stars form and die and
in the process they increase the absolute amount of iron in the ISM.
Clearly, if the star formation rate is lower, the rate of decline of
[Mg/Fe] is lower and the initial high value of this ratio is
maintained longer, as shown in panels c). In our models the star
formation rate is higher in the inner regions than in the outer disk
as shown in panel (b) of Figures 7,8 and 9; therefore the ratio
[Mg/Fe] decreases rapidly in the inner disk, and maintains a larger
value at larger radius.  As a result, the (positive) radial gradient
for the [Mg/Fe] ratio is larger at present than at early times, as
seen in panel (d). Notice that the radial distributions of gas
abundances are shown for four different ages, 2.5, 5, 8.5 and 13 Gyr.

In this section we have concentrated so far on the abundance of the
gas. Let's consider now star formation. Stellar populations must have
the same abundances the gas had at the time these stars formed from
it. We must keep in mind that stars are continuously forming in spiral
disks, and that the star formation history is different at different
galactocentric radii as already discussed.  The inner regions of disks
had formed a large number of stars during the first 3 Gyr and
subsequently the rate declined abruptly, while in the periphery the
number of stars formed now is similar to that of the last 8
Gyr. Therefore the stellar populations of the central regions were
formed when iron abundances were still as low as [Fe/H]=-0.4, whereas
[O/H] and [Mg/H] had already reached almost solar values.  The outer
regions, on the other hand, are forming stars even now and the [Mg/Fe]
ratio has decreased as a result of the iron abundance accumulated in
the ISM being large by now. We conclude that the spatial distribution
of abundances for the stellar populations must be flatter than that of
the gas. This is consistent with what we show in panel (d) of the same
figures.  The ongoing discussion reinforces the conclusion that the
radial behavior of the [Mg/Fe] ratio depends strongly on the SFR and
thus that our attempt at extracting this ratio from the integrated
spectral indices of the stellar disk should produce a strong
constraint to the models.

The preceding discussion shows why at the present time the stellar
radial gradient in the mean iron abundance of the population, $<[\rm
Fe/H]>$, is smaller than the corresponding one for the gas. The
stellar radial gradient of magnesium, on the other hand, has almost
the same value as the present gradient of the gas abundance. This
result is clearly seen in the case of NGC~ 4321, where the stellar
mean iron abundance can be fitted with a straight line with a positive
slope ($\sim +0.003$ dex/kpc), while the stellar magnesium abundance
has an almost (slightly negative) flat gradient ($\sim$ -0.0014
dex/kpc).  The final result of evolution is a flattening in the iron
index radial distribution and a steep radial gradient in the magnesium
index Mg$_{2}$.

Finally, we wish to add some comments on the overall time evolution of
each of our program galaxies. We can see that NGC~ 4321 is a more
evolved galaxy than NGC~ 4303 which is still forming stars.  NGC ~
4535 has an intermediate behavior. It would therefore seem that the
time evolution is determinated to a large extent by the total mass of
each galaxy: NGC~ 4303 is the smaller, and therefore collapsed slower
than the two other while maintaining the highest surface density of
atomic gas at the maximum of the radial distribution.  Moreover, as
discussed in MFD97, this maximum moves outwards across the disk as
time progresses and it is found at larger radius for more evolved
galaxies. Following this argument we conclude that NGC~ 4303 is the
less evolved of the three, whereas galaxies NGC~ 4321 and NGC~ 4535
are essentially at the same stage of evolution with a maximum located
between 10 and 12 kpc. This is also shown by the molecular gas
densities. In effect, NGC~ 4303 has the highest surface density, with
NGC~ 4535 and NGC~ 4321 having lower central values. Thus NGC~ 4303
may still be able to consume a large quantity of molecular gas to form
stars.

\section{CONCLUSIONS}

We have measured for the first time here and in Paper I integrated
spectral indices over the disks of galaxies. These indices were used
 as additional constraints to the multiphase chemical evolution models.
We have therefore added information on the past evolution of the gas
and stars to that derived from the study of H\,{\sc ii} regions, which
belong to the youngest population. The latter provide a chemical
composition in $\alpha$ elements which is the accumulated result of the
chemical evolution of the galaxy over its entire history. Our indices,
on the other hand, are related to a mean abundance weighted by the SFR,
and correspond to an older mean age than that of H\,{\sc ii} regions.
Therefore, they provide the models with a different point in time, but
at the expense of a complex strongly model-dependent transformation of
indices into abundances. The extraction of abundances from H\,{\sc ii},
on the other hand, is a comparatively simpler procedure which depends
on reasonably well understood physics, but that can only provide the
present-day abundance of the gas.

The multiphase chemical evolution model as implemented here fits
radial distributions of diffuse gas densities quite well.  Both
molecular gas and star formation rate surface density are also well
reproduced after assuming that star formation from cloud--cloud
collisions has a radial dependence due to the effect of the velocity
dispersion or, equivalently, that star formation is possible only
after a threshold gas density is reached. Radial distributions of
abundances show  reasonable agreement with the data. However, the
high oxygen abundances observed, mostly in the central regions, cannot
be reproduced by our models. Probably, radial flows and/or gas lost
must be included in order to reproduce the abundance observed in our
Virgo cluster test galaxies.

We find that the abundance of the gas is larger in the inner regions
of disks, but that the ages of the stellar populations created from
this gas are different, being younger in the external regions than
in the inner disk. In order to explain the radial gradients
found from the absorption spectral index distributions, both the age
and the metallicities of the stellar disk must be taken into account.

The abundances of both the gas and the stars are larger in the inner
regions because there the star formation rate has a maximum which is
both larger and earlier than the equivalent maximum of the outer
regions. Stellar abundances have a flatter radial distribution than
that of the gas at the present time, but this flattening is stronger
for the iron than for the magnesium. As a result, the radial
distributions of the magnesium feature are steeper than those of the
iron index.

The effect of the environment on the chemical evolution must
 eventually be taken into account. It is not clear whether chemical
 abundances are affected by it (Shields, Skillman, \& Kennicutt 1991;
 Henry 1993; Henry et al. 1992,1996; Henry, pagel, \& Chincarini
 1994).  It seems clear however that there is a H\,{\sc i} deficiency
 in cluster galaxies (Giovanelli \& Haynes 1985) and that the radial
 distributions of diffuse gas seem more concentrated as if the gas was
 lost from the outer regions of disks.

Finally, an important parameter, not specifically included 
in the chemical evolution models described here is the existence of
bars.  Barred galaxies have a flatter radial gradient (Edmunds \& Roy
1993; Martin \& Roy 1992,1994,1995).  It has been suggested (Friedli,
Benz, \& Kennicutt 1994; Friedli \& Benz 1995; Roy 1996) that bars
induce large--scale mixing as a result of radial flows.  These radial
flows produce a large accumulation of mass in the inner regions of the
disk and, as a result, a burst of star formation. The final outcome is
a flatter abundance gradient in the outer disk and a steepening of
this gradient in the inner disk.  It should be stressed that the
multiphase model uses as input the total mass radial distribution,
obtained from the rotation curve. But the possibility of a rotation
curve changing with time as a result of dynamical processes such
as radial flows of gas are not included in this model. This issue, as
well as a study of the sensitivity of the models to the input
parameters remains to be studied, and will be the subject of the next
papers in this series.

\begin{acknowledgements}

M.Moll\'{a} acknowledges the Spanish {\sl Ministerio de Educaci\'{o}n y
Cultura} for its support through a post-doctoral fellowship and to the
astrophysics group of the Universit Laval for the nice work atmosphere
during her stay in this group.  E. Hardy acknowledges partial support
from NSERC-Canada and from FCAR-Qu\'ebec during the early stages of
this investigation.  We thank the referee, Daniel Friedli, for his
useful comments.  This research has made use of the
NASA/IPAC Extra-galactic Database (NED), which is operated by the Jet
Propulsion Laboratory, Caltech, under contract with the National
Aeronautics and Space Administration.

\end{acknowledgements}

\newpage
\noindent

\centerline{FIGURE CAPTIONS}

Figure 1. Predicted present--time radial distributions for NGC~ 4303:
a) Oxygen abundance $12+log (O/H)$; b) Star Formation Rate surface
density normalized to the `equivalent to solar region value'
$\Psi/\Psi_{0}$; c) diffuse or atomic gas H\,{\sc i} and d) molecular
gas H$_{2}$ surface densities.  Model is represented by the solid
line.  Open symbols are the observational data.

Figure 2. The same as Figure 1 for NGC~ 4321.

Figure 3. The same as Figure 1 for NGC~ 4535.

Figure 4. Radial distributions of spectral indices for NGC~ 4303: a)
Mg$_{2}$, b) Fe5270.  In both cases the predicted radial distributions
are represented by the solid line. Open circles with error bars are
observational data.

Figure 5. The same as Figure 4 for NGC~ 4321.

Figure 6. The same as Figure 4 for NGC~ 4535.

Figure 7. Time evolution for three different regions located in the
inner, the medium and the outer disk of NGC~ 4303: a) magnesium (solid
line) and iron (dashed line) abundances as $12+log (X/H)$; b) The star
formation history $ log {\Psi/\Psi_{0}} $, c) the relative abundance
[Mg/Fe], d) the radial gradient of the relative abundance [Mg/Fe] for
different times: the dotted line is the result for $\rm t=2.5$ Gyr,
the dot-dashed line for $\rm t=5$ Gyr, and the dashed line for $\rm
t=8.5$ Gyr. The present radial distribution is represented by the
dot-long dashed line. The solid line is the mean stellar abundance.

Figure 8. The same as Figure 7 for NGC~ 4321.

Figure 9. The same as Figure 7 for NGC~ 4535.

\newpage


\begin{table*}
\caption{Filter Characteristics}

\begin{tabular}{lcc}
& &  \\
\tableline\tableline
 Filter & $\lambda _\circ$ & $\Delta \lambda$ \\
 &            	[\AA ]	&		[\AA ]\\
\tableline
BC &		4964 &			59 \\
Mg$_{2}$	&	5214 &			56 \\
Fe52	&	5300 &			55 \\
RC	&	5376 & 			59 \\
\tableline

\end{tabular}
\end{table*}

\begin{table*}
\caption{ Galaxy Characteristics}
\begin{tabular}{lcccccc}
& & & & & \\
\tableline\tableline
\noalign{\smallskip}
Name & Type & Arm Class & D (Mpc)&  $R_{eff}$ (arcsec)
& $V_{ max}$(km/s) \\
\tableline
NGC~ 4303 & 4 & 9 & 15.2 & 62.6  & 150 \\
NGC~ 4321 & 4 & 12 & 16.8 & 111.5 & 270 \\
NGC~ 4535 & 5 & 9 & 16.8 &  104  & 210 \\
\tableline
\end{tabular}
\end{table*}

\clearpage
\begin{table*}
\caption{Model Input Parameters}
\begin{tabular}{lcccc}
& & & & \\
\tableline\tableline
Galaxy & $R_{0}$  & $\tau (R_{0})$ 
&$\epsilon_{\mu}$ & $\epsilon_{ H}$\\
Name &  (kpc) & (Gyr) & & \\
\tableline
NGC~ 4303 &6 & 8 & 0.22 & 0.015 \\
NGC~ 4321 &9 & 5 & 0.45 & 0.10 \\
NGC~ 4535 &8 & 5 & 0.28 & 0.02 \\ 
\tableline
\end{tabular}
\end{table*}

\end{document}